\documentclass[11pt,a4paper]{article}
\usepackage{jheppub}
\usepackage{graphicx}
\usepackage{rotating}
\usepackage{color}
\usepackage{epsfig}
\usepackage{bm}
\usepackage{slashed}
\usepackage{array}

\newcommand{\be}{\begin{equation}}
\newcommand{\ee}{\end{equation}}
\newcommand{\bea}{\begin{eqnarray}}
\newcommand{\eea}{\end{eqnarray}}
\newcommand{\bi}{\begin{itemize}}
\newcommand{\ei}{\end{itemize}}
\newcommand{\ben}{\begin{enumerate}}
\newcommand{\een}{\end{enumerate}}
\newcommand{\bt}{\begin{tabular}}
\newcommand{\et}{\end{tabular}}

\newcommand{\lc}{\left[}
\newcommand{\rc}{\right]}
\newcommand{\lp}{\left(}
\newcommand{\rp}{\right)}

\newcommand{\op}{\mathcal{O}}
\newcommand{\Opann}{\op_{\rm ann}}

\newcommand{\timeord}{\mathcal{T}}
\newcommand{\fig}{Fig.~}

\newcommand{\unitOP}{1\!\!1}

\def\sl{\slashed }

\newcommand{\bra}[1]{\langle #1|}
\newcommand{\ket}[1]{|#1\rangle}

\def\Tr{{\rm{Tr}}}

\def\Im{{\rm{Im}\,}}

\newcounter{MBQ}

\def\c{\chi }

\def\lg{{\mathchoice{~\raise.58ex\hbox{$<$}\mkern-14.8mu\lower.52ex\hbox{$>$}~}
                    {~\raise.58ex\hbox{$<$}\mkern-14.8mu\lower.52ex\hbox{$>$}~}
                    {\raise.59ex\hbox{{$\scriptscriptstyle <$}}\mkern-12.8mu%
                     \lower.01ex\hbox{{$\scriptscriptstyle >$}}}   {}   }} 
\def\gl{{\mathchoice{~\raise.58ex\hbox{$>$}\mkern-12.8mu\lower.52ex\hbox{$<$}~}
                    {~\raise.58ex\hbox{$>$}\mkern-12.8mu\lower.52ex\hbox{$<$}~}
                    {\raise.62ex\hbox{{$\scriptscriptstyle >$}}\mkern-12.0mu%
                     \lower.05ex\hbox{{$\scriptscriptstyle <$}}}  {}    }}  

\allowdisplaybreaks

\begin{document}

\begin{flushright}
TUM-HEP 1052/16\\
%arXiv:1607.0nnnn [hep-ph]\\
July 12, 2016
\end{flushright}

\vskip 0.2in

\title{Finite-temperature modification of heavy particle decay 
and dark matter annihilation}

\author[a]{Martin Beneke,}
\author[a]{Francesco Dighera,}
\author[b,c]{Andrzej Hryczuk}

\affiliation[a]{Physik Department T31, Technische Universit\"{a}t 
M\"{u}nchen, James-Franck-Str.~1, 85748 Garching, Germany}
\affiliation[b]{Department of Physics, University of Oslo, Box 1048, 
NO-0371 Oslo, Norway}
\affiliation[c]{National Centre for Nuclear Research, Ho\.za 69, 00-681, 
Warsaw, Poland}
\emailAdd{francesco.dighera@tum.de}
\emailAdd{a.j.hryczuk@fys.uio.no}

\date{\today}

\abstract{
We apply the operator product expansion (OPE) technique 
to the decay and annihilation of heavy particles in a thermal medium 
with temperature below the heavy particle mass, $m_\chi$. This allows us to 
explain two interesting observations made before: a) that the leading 
thermal correction to the decay width of a charged particle is the 
same multiplicative factor of the zero-temperature width for 
a two-body decay and muon decay, and b) that the 
leading thermal correction to fermionic dark matter annihilation arises 
only at order $T^4/m_\chi^4$. The OPE further considerably simplifies the 
computation and factorizes it into model-independent matrix elements 
in the thermal background, and short-distance coefficients to be 
computed in zero-temperature field theory. 
}

\maketitle

%%%%%%%%%%%%%%%%%%%%%%%%%%%%%%%%%%%%%%%%%%%%%%%%%%%%%%%%%%%%%%%%%%%%%%%%%%%%

\vspace*{0.2cm}
\section{Introduction}
\label{sec:Intro}

In the early Universe all interactions between particles take place in 
a thermal plasma, leading to modifications of the cross sections and 
corresponding rates with respect to their zero-temperature values. 
Many interesting phenomena occur in the regime where the temperature $T$ 
is much smaller than the energy scale $M$ of the hard process. Examples 
are the freeze-out of weakly interacting massive dark matter particles and 
the decay of heavy particles during big-bang nucleosynthesis. In these 
cases the finite-temperature modification is expected to be of order  
$g^2\times (T/M)^a$, where $g$ represents the interaction strength and 
$a$ is some number.

Modifications to particle decay widths due to a thermal bath of photons 
have been studied in different cases and remarkably simple results were 
found. In particular, the leading $\mathcal{O}(T^2)$ correction vanishes 
for the decay of the neutral Higgs boson~\cite{Donoghue:1983qx} and heavy 
Majorana neutrino~\cite{Biondini:2013xua}, while for a charged particle 
two-body decay it is simply proportional to the tree level 
width~\cite{Czarnecki:2011mr}. Interestingly, the same result was found for 
the three-body decay of the muon. The more involved case of annihilation 
of two neutral fermions $\chi$ into two charged fermions via $t$-channel 
exchange of a massive scalar has been computed in~\cite{Beneke:2014gla}. 
The primary purpose of that work was to demonstrate the infrared divergence 
cancellation in relic density computations, but the leading 
temperature-dependent correction was also considered and found to appear 
only at $\op(T^4)$.

The simplicity of these results calls for an explanation, which we provide 
here in the framework of the operator product expansion (OPE). In the 
case of particle decay this treatment is inspired by, and in many ways 
very similar to, the OPE of inclusive heavy quark decay in 
QCD~\cite{Bigi:1992su}. The thermal background plasma takes the role of the 
soft QCD vacuum quantum fluctuations. 
For dark matter annihilation we apply a generalized 
operator production expansion to the (imaginary part of the) 
$\chi\chi\to\chi\chi$ forward scattering-amplitude.
In addition to providing a physical understanding of the observed 
universality and simplicity of the $\mathcal{O}(T^2)$ corrections, the
OPE offers a considerable simplification of the calculations.  
The approach presented here can be straightforwardly applied to other 
physically motivated situations such as the co-annihilation of 
charged states during freeze-out. We note that OPE methods have been 
developed systematically before for thermal field theory in the QCD context 
for SVZ sum rules at finite temperature 
\cite{Hatsuda:1992bv,Mallik:1997pq}, and for the study of more 
general spectral functions in the low-energy QCD plasma 
\cite{CaronHuot:2009ns}. 

%%%%%%%%%%%%%%%%%%%%%%%%%%%%%%%%%%%%%%%%%%%%%%%%%%%%%%%%%%%%%%%%%%%%%%%%%%%%

\section{Charged particle decay at finite temperature}
\label{sec:decay}

For definiteness we consider the spin-averaged total width of a 
fermion $\psi$ with electric charge $q$ into another fermion $\chi$ of the 
same electric charge and other neutral particles in an unpolarized thermal 
bath of photons and SM fermions $f$ at a temperature $T$. Specific examples 
are those of~\cite{Czarnecki:2011mr}, $\psi\to \chi\phi$, where $\phi$ 
is a neutral scalar, and the more realistic muon decay $\mu \to e 
\nu_\mu\bar\nu_e$. We assume that the temperature of the bath is small 
compared to the $\psi$ mass,  $m_\psi \gg T$, but can be of the same order 
or even much larger than mass of the other charged particle.
We further assume that the decay occurs at rest with respect to the thermal 
bath. The decay rate will be modified compared to the zero-temperature 
value by interactions with the plasma. 

The decay of $\psi$ is caused by a weak interaction
\begin{equation}
{\cal L} = \lambda \,J_\mu \,{\cal O}^\mu + {\rm h.c.}\,
\end{equation}
where $J_\mu$ is the fermion current and ${\cal O}^\mu$ represents the 
neutral fields. For the toy situation $\psi\to \chi\phi$, 
$J_\mu = \bar\chi P_L\psi$ ($P_L=\frac{1-\gamma_5}{2}$) 
and ${\cal O}_\mu=\phi$, while for muon decay 
$J_\mu = [\bar e \mu]_{\rm V-A}$, ${\cal O}_\mu = [\bar 
\nu_\mu\nu_e]_{\rm V-A}$ and $\lambda=-G_F/\sqrt{2}$. By the optical 
theorem the decay width can be expressed as 
\be
\label{eq:gamma_def}
\Gamma_T = \lambda^2 L^{\mu\nu}  \, 
2 \,\textrm{Im}\left\{ T_{\mu\nu} \right\} 
\ee
to lowest order in the weak coupling $\lambda$, but to all orders 
in the electromagnetic interaction. Here 
$\textrm{Im}\left\{ T_{\mu\nu} \right\}$ refers to the imaginary part of 
\be
T_{\mu\nu}= \frac{1}{2}\sum_\textrm{spin}\, (-i)\int d^4x\ e^{-ip\cdot x}\ 
\bra{\psi;T}\ \timeord\{J_\mu(0)\, J_\nu^\dagger(x)\} \ket{\psi;T},
\label{eq:Tmunu}
\ee
and $L^{\mu\nu}$ to the neutral particles, which are unaffected by 
the plasma, integrated over phase space. $\ket{\psi;T}$ denotes the 
$\psi$ one-particle state with momentum $p=m_{\psi;T} v$ and 
non-relativistic normalization in the thermal 
bath. We can decompose $p^\mu=m_\psi v^\mu + k^\mu$, where $v$ is the 
four-velocity of the plasma and $m_\psi$ the zero-temperature mass. Since we 
assume that the decaying particle is at rest with respect 
to the plasma, and $T\ll m_\psi$, $k$ is a soft momentum with scaling 
$k\sim T$. 

The scale hierarchy $T\ll m_\psi$ allows us to separate the hard decay 
process from the effects of the thermal bath by performing the OPE 
of the correlation function (\ref{eq:Tmunu}). The short-distance physics is 
encoded in the Wilson coefficients, which can be computed at zero 
temperature, while the thermal modifications all reside in the matrix 
elements of local operators computed in the thermal bath.
The situation is analogous to the calculation of the (zero-temperature) 
semi-leptonic decay width \cite{Chay:1990da,Blok:1993va, Manohar:1993qn}
of a heavy $b$-hadron $H_b$. The $\psi$ particle plays 
the role of the $b$ quark, while the thermal bath substitutes the hadronic 
medium of soft light quarks and gluons in $H_b$. The finite-temperature 
$\psi$ mass $m_{\psi;T}$ is the analogue of the B-hadron mass, the 
zero-temperature mass corresponds to the quark mass.

The relevant OPE of the time-ordered product in (\ref{eq:Tmunu}) is 
\be
-i\int d^4x\ e^{-ip\cdot x}\ \timeord\{J^\mu(0)\, J^{\nu\dagger}(x)\} = 
C^{\mu\nu}_0 \bar\psi \psi + 
C^{\mu\nu}_2 \bar\psi \frac{i}{2}\sigma_{\alpha\beta}F^{\alpha\beta} \psi + 
\mathcal{O}(m_{\psi}^{-3}),
\label{eq:OPE}
\ee
where the Wilson coefficients $C^{\mu\nu}_0$ and $C^{\mu\nu}_2$ are specific 
to the particular decay process. However, they depend only on $m_\psi v$, 
and are to be computed by matching at $T=0$. Hence only the zero-temperature 
mass $m_\psi$ enters. In general, the background plasma breaks Lorentz 
invariance and it might be necessary to keep additional operators, which 
are not scalars, and have non-vanishing matrix elements in the thermal 
bath. However, since we assume that the particle decays at rest with 
respect to the plasma, the vector $v$ coincides with the one 
already introduced by the particle state itself, and (\ref{eq:OPE}) 
for muon decay is the same as appears in \cite{Bigi:1993fe}. 
The neglected term can contribute at most at ${\cal O}(T^3/m_\psi^3)$, 
which is smaller than the putative leading thermal correction.

As the second (magnetic) operator does not contribute in an unpolarized 
medium, it remains to evaluate the matrix element $\bra{\psi;T} 
\,\bar\psi \psi\, \ket{\psi;T}$ of the leading operator in the thermal 
background. Since this is independent of the short-distance decay, 
it already follows here that the thermal correction must be a universal 
modification of the zero-temperature decay width. The matrix elements 
of the OPE in heavy particle states have themselves a non-trivial 
$1/m_\psi$ expansion. Using the equation of motion, we can write 
\cite{Blok:1993va}
\be
\bar \psi\psi= \bar\psi \slashed v \psi +\frac{1}{2m_\psi^2} \bar\psi 
\left(iD_\perp\right)^2 \psi + 
\frac{i}{4m_\psi^2}\bar\psi \sigma_{\alpha\beta}F^{\alpha\beta} \psi + 
\mathcal{O}(m_{\psi}^{-3}),
\ee
where the transverse covariant derivative is defined as 
$D_\perp^\mu \, \equiv \, g_\perp^{\mu\nu} D_\nu\, \equiv\, \lp  g^{\mu\nu} 
- v^\mu v^\nu \rp D_\nu$. The usefulness of this equation stems from the 
fact that the first term is related to a conserved current, and hence 
is matrix element is known exactly, 
\be
\label{eq:M_gammamu_full}
\bra{\psi;T} \, \bar\psi \gamma^\mu \psi \, \ket{\psi;T} = v^\mu
\ee
from which it follows that
\be
T^{\mu\nu}= C^{\mu\nu}_0 \,\bigg(1 + \frac{1}{2m_\psi^2} \bra{\psi;T}  \,
\bar\psi \left(iD_\perp\right)^2 \psi \,\ket{\psi;T} \bigg) 
+\mathcal{O}(m_{\psi}^{-3}).
\ee
Therefore the leading thermal correction is a direct effect of the matrix 
element of the kinetic energy operator $\mathcal{O}_k\equiv -\bar\psi 
\frac{\left(iD_\perp\right)^2}{2m_\psi^2} \psi $ evaluated in the thermal 
background. We may interpret $K_\psi = \bra{\psi;T}  
\mathcal{O}_k \ket{\psi;T}$ as the average kinetic energy of the 
$\psi$ particle acquired by the interactions with the photons in the 
plasma. By dimensional analysis $K_\psi ={\cal O}(T^2/m_\psi^2)$.
Since $\Gamma_0 \equiv 
\lambda^2 L_{\mu\nu}  \, 2 \,\textrm{Im}\left\{ C_0^{\mu\nu} \right\}$ 
is the zero-temperature width, the decay width at finite temperature 
reads
\be
\Gamma_T = \Gamma_0 \left(1 - K_\psi \right) +\mathcal{O}(T^{3}/m_\psi^3). 
\label{eq:GammaT}
\ee
This derivation explains in a rather straightforward manner three 
observations originally made in~\cite{Czarnecki:2011mr}: 1) that soft 
and collinear divergences cancel in the sum of virtual corrections and 
emission and absorption processes, 2) the leading finite-temperature 
correction is ${\cal O}(T^2/m_\psi^2)$, and 3) a universal 
factor multiplying the tree-level decay width.\footnote{It is worth 
noting that at $\mathcal{O}(T^4)$, more operators contribute and the 
matching coefficients depend on the details of the hard process. 
Nevertheless, the temperature-dependent part arises from matrix 
elements of local operators and the OPE greatly simplifies the calculation 
of such sub-leading terms.}

%%%%%%%%%%%%%%%%%%%%%%%%%%%%%%%%%%%%%%%%%%%%%%%%%%%%%%%%%%%%%%%%%%%%%%%%%%%
\begin{figure}[t] 
 \centering
 \includegraphics[width=0.8\textwidth]{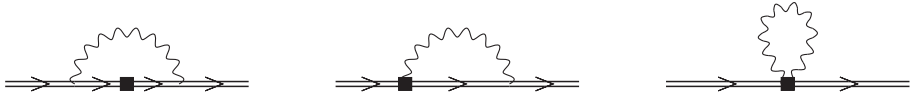}
 \caption{The diagrams defining the one-loop contributions 
$K_1$, $K_2$ and $K_3$, respectively, to the kinetic operator matrix 
element. The filled square represents the operator insertion and 
diagrams 1 and 2 involve the usual QED vertices aside from the 
operator vertex.}
 \label{fig:diags_muon}
\end{figure}
%%%%%%%%%%%%%%%%%%%%%%%%%%%%%%%%%%%%%%%%%%%%%%%%%%%%%%%%%%%%%%%%%%%%%%%%%%%

In contrast to the QCD analogue of the semi-leptonic decay of $H_b$, where 
the soft physics is non-perturbative, the matrix element $K_\psi$ of the 
kinetic operator in the thermal plasma can easily be computed 
perturbatively. The one-loop diagrams are depicted in 
\fig\ref{fig:diags_muon}, in terms of which the matrix element is given 
by the sum $K_\psi =K_1+ 2 K_2 + K_3$.
Since we are interested in the temperature-dependent correction, 
we retain only the thermal part of the equilibrium photon propagator
\bea
iD^{11}_{\mu\nu}(x,y) &=& 
\bra{\Omega;T}\ \mathcal{T}\{ A_\mu(x) A_\nu(y)\}\ \ket{\Omega;T} 
\nonumber \\
&=& \int\frac{d^4k}{(2\pi)^4}\ e^{-ik\cdot(x-y)}\ (-2\pi)g_{\mu\nu} 
\delta(k^2) \, f_B(k^0),
\label{eq:Tpropgamma}
\eea
where $f_B(k^0) = (e^{|k^0|/T}-1)^{-1}$ is the Bose-Einstein distribution 
of the photons in the rest frame of the plasma. Writing down the diagrams 
explicitly, we find the spin-averaged matrix elements 
\bea
K_1 & = & \frac{i(ie)^2}{2m_\psi^2}  \int\frac{d^4k}{(2\pi)^4}\, 
iD^{11}_{\alpha\beta}(k)\,
\frac{1}{4 m_\psi}\,\mbox{tr}\bigg[(\slashed p+m_\psi)
\gamma^\alpha
\frac{1}{\slashed p - \slashed k - m_\psi} 
\nonumber\\
&&\hspace*{1.5cm} (-i)\,(p_\perp \! -k_\perp)^2 
\gamma^\beta\frac{1}{\slashed p - \slashed k - m_\psi}\bigg], 
\\[0.1cm]
K_2 & = & \frac{i(ie)^2}{2m_\psi^2}  \int\frac{d^4k}{(2\pi)^4}\, 
iD^{11}_{\alpha\beta}(k)\,(2p_\perp + k_\perp)^\beta 
\frac{1}{4 m_\psi}\,\mbox{tr}\left[(\slashed p+m_\psi)
\gamma^\alpha\frac{1}{\slashed p - \slashed k - m_\psi}\right], 
\quad\label{eq:K_mu}\\[0.1cm]
K_3 & = & \frac{i(ie)^2}{2m_\psi^2} \int\frac{d^4k}{(2\pi)^4}\, 
iD^{11}_{\alpha\beta}(k)\, (-ig_\perp^{\alpha\beta}) \,
\frac{1}{4 m_\psi}\,\mbox{tr}\left[\slashed p+m_\psi\right].
\eea
Note that the particle mass in the thermal background, $m_{\psi, T}$, 
rather than the Lagrangian mass $m_\psi$ should be used in the 
evaluation of the low-energy matrix elements, but since the difference 
of the mass-squares is ${\cal O}(T^2)$, it contributes to a higher-order 
correction, and both can be identified. The above expressions 
can be simplified by using 
$p=m_\psi v$ with $v^2=1$, and $v\cdot k_\perp=0$. The straightforward 
evaluation then results in 
\be
K_\psi  =  -\frac{\pi}{6}\alpha \tau^2 + 2\times 0 +  
3\, \frac{\pi}{6}\alpha \tau^2 + \op\lp \tau^3\rp = 
\frac{\pi}{3}\alpha \tau^2 + \op\lp \tau^3\rp,
\ee
where we defined $\tau=T/m_\psi$. This has to be multiplied by $q^2$ 
for a particle with electric charge $q$ in units of the positron charge.
Inserting this into (\ref{eq:GammaT}) gives, explicitly, 
\be
\label{eq:gamma_gen}
\Gamma_T = \Gamma_0 \lp 1 - \frac{\pi}{3}\alpha q^2 \tau^2 \rp + 
\op\lp\tau^3\rp\ ,
\ee
in complete agreement with~\cite{Czarnecki:2011mr}. 

Comparison with the derivation of this result in~\cite{Czarnecki:2011mr}
highlights the power of the OPE approach. It also provides a physical 
interpretation of the correction as a time dilatation effect due to the 
average kinetic energy of the particles due to collisions with 
the photons of the plasma.

The finite-temperature modification of the decay width of a Majorana 
fermion~\cite{Biondini:2013xua} mentioned in the introduction was obtained 
through effective field theory methods, which are also based on 
systematic scale separation. The OPE nevertheless provides a more 
direct approach to the inclusive decay width in the same way as 
the full development of heavy quark effective theory is not required 
to compute the $1/M$ expansion of the inclusive or semi-leptonic 
$b$-hadron decay width in QCD. 

%%%%%%%%%%%%%%%%%%%%%%%%%%%%%%%%%%%%%%%%%%%%%%%%%%%%%%%%%%%%%%%%%%%%%%%%%%%%

\section{Dark matter annihilation}
\label{sec:AnnihilT4}

Effective field theory and the OPE can also be applied to two-particle 
annihilation in the thermal medium, provided the temperature is small 
compared to the annihilating particles' mass. This considerably 
simplifies the diagrammatic analysis of~\cite{Beneke:2014gla} and provides 
an understanding of the temperature scaling of the leading thermal correction. 
Although the method is general, we discuss below the annihilation of a
heavy, electrically neutral Dirac fermion into a pair of light charged 
fermions ($m_f\ll T$), 
and refer to the heavy fermion as the ``dark matter'' particle  
to establish contact with~\cite{Beneke:2014gla}. 
 
We assume at first that the annihilation process $\chi\bar\chi\rightarrow 
f\bar f$ occurs through the local four-fermion operator 
\begin{equation}
\Opann=\frac{1}{\Lambda^2} \,(\bar\chi \,\Gamma^A \chi)\,
( \bar f \,\Gamma_A' f)\,,
\label{eq:Oann}
\end{equation}
where $1/\Lambda^2$ is an unspecified coefficient of mass dimension $-2$, and 
$\Gamma^A$, $\Gamma_A^\prime$ are Dirac matrices, which may be multiplied 
by up to one covariant derivative. 

The total spin-averaged annihilation cross section in the 
thermal background follows from the optical theorem, 
\bea
\sigma_{\rm ann} \,v_{\rm rel}  &=& \frac{2}{s}\;\Im \,\bigg\{(-i)  
\int d^4x\,\frac{1}{4}\,\sum_{\rm spin} \bra{\bar\chi\chi;T}\,  
\timeord \left\{ \Opann(0)\, \Opann^\dagger(x) \right\} \,
\ket{\bar\chi\chi;T}  \bigg\}\, ,
\label{eq:sigmav}
\eea
where the state $\ket{\bar\chi\chi;T}$ represents the annihilating pair 
in the thermal photon background, $p\equiv p_1+p_2$ is the total incoming 
momentum and $s\equiv p^2$ the center-of-mass energy squared. Once again, 
we assume that the center-of-mass frame of the annihilation is at rest 
with respect to the plasma, in which case  $p = \sqrt{s}\,v$ with $v$ 
defining the plasma frame. 

Since the annihilating particles do not couple to the thermal bath, 
the $\chi$ field part of (\ref{eq:sigmav}) is readily done by 
contracting the fields with the external state, which results 
in some tensor $L_{AB}$. The non-trivial part is the time-ordered 
product of the fermion current $J_A =  \bar f \,\Gamma^\prime_A f$. 
Since the final state particles are very energetic relative to 
the soft degrees of freedom of the plasma, we perform the OPE 
\be
-i \int d^4x\ e^{-ip\cdot x} \ \timeord \left\{J_A(0)\, 
J_B^{\dagger}(x) \right\}\ = \sum_{i} C_{AB}^i(p)\cdot \op_i\ ,
\label{eq:OPE2}
\ee
of which the matrix element within the thermal vacuum $|\Omega_T\rangle$ 
needs to be taken. Up to dimension 4, the operators $\op_i$ are constructed 
from contractions with the metric tensor and the plasma velocity $v$ of 
\be
\label{eq:ops_with_indices}
\unitOP \ ,\qquad F^{\alpha\beta}F^{\gamma\delta} \ , \qquad 
m_f \bar f\, \Gamma f\ ,\qquad \bar f\, \Gamma\, iD^\alpha f\ ,
\ee
with $\Gamma$ a general Dirac matrix in spinor space. Apart from the unit 
operator there is no operator of dimension lower than 4. This allows us 
to deduce immediately that the leading order thermal correction is at least 
of the order $\op(T^4)$ or $\op(m_f^2 T^2)$.

The thermal matrix elements are easily computed, see Appendix~\ref{sec:AppA}. 
The photon ``condensate'' is given by 
\begin{eqnarray}
\bra{\Omega_T} \, F^{\alpha\beta}F^{\gamma\delta} \, 
\ket{\Omega_T} &=& \frac{\pi^2}{45}\,T^4 \,\Big\{
\lp g^{\alpha\gamma}g^{\beta\delta} - g^{\alpha\delta}g^{\beta\gamma} \rp
\nonumber\\[0.1cm]
&& -\,2  \lp v^{\alpha } \big(v^{\gamma } g^{\beta  \delta }-v^{\delta } 
g^{\beta  \gamma }\big) - 
v^{\beta } \big(v^{\gamma } g^{\alpha  \delta }- v^{\delta } 
g^{\alpha  \gamma }\big) \rp\Big\}\,.
\quad 
\label{eq:photoncondensate}
\end{eqnarray}
The light fermion operator $m_f \bar f\, \Gamma f$ is generated only 
with $\Gamma = \unitOP$ due to parity invariance and helicity conservation 
and is power-suppressed for $m_f \ll T$ (and exponentially suppressed 
for $m_f \gg T$, but we do not consider this limit): 
\be
\bra{\Omega_T}\,m_f \bar f f\, \ket{\Omega_T} = 
\mathcal{O}(m_f^2 T^2) \ll \mathcal{O}(T^4)\,.
\ee
The addition of a covariant derivative alleviates the $m_f$ suppression.  
In the limit $m_f\rightarrow 0$ we obtain
\be
\bra{\Omega_T}\, \bar f\, \gamma^\mu iD^\alpha f\, \ket{\Omega_T} 
=  - \frac{7\pi^2}{180} \,T^4 \,\Big(\,g^{\mu\alpha} - 
4\, v^\mu v^\alpha\Big)\,.
\label{eq:fermioncondensate}
\ee

The short-distance coefficients from contracting the $\chi$ fields and 
the OPE depend on the momenta $p_1$ and $p_2$ of the two annihilating 
particles. Since $p_1+p_2=p=\sqrt{s} v$, 
\bea
\label{eq:O_Afinal}
\op_{A\,1}\equiv \lp p_1^\mu p_2^\nu F_{\mu\nu} \rp^2, \quad
\op_{A\,n}\equiv p_i^\mu p_j^\nu F_{\alpha\mu} F^\alpha_{\ \nu} \ , \quad
\op_{f\, n} \equiv \bar f\, \slashed p_i\,  p_j^\mu \,iD_\mu \,f 
\eea
for $n\equiv i+j=2,3,4$ form a complete set of scalar operators. Using 
(\ref{eq:photoncondensate}), (\ref{eq:fermioncondensate}) their matrix 
elements in the state $|\Omega_T\rangle$ are given by 
\bea
\label{eq:M_Afinal}
M_{A\, 1} &=&  m_\chi^4\,\frac{\pi^2}{45} \,T^4 \, 
4e_\chi^2 \lp e_\chi^2-1 \rp , \quad\qquad \\
M_{A\, 2} = M_{A\, 4} &=& m_\chi^2 \,\frac{\pi^2}{45} \,T^4 \,
(-1) \lp 4e_\chi^2-1 \rp , \\
M_{A\, 3} &=& m_\chi^2 \,\frac{\pi^2}{45} \,T^4 \, 
(-1)\lp 2e_\chi^2+1 \rp , \\
M_{fn} &=& -\frac{7}{4} M_{A\, n}\qquad (n\not=1)\,,
\eea
with $e_\chi=E_\chi/m_\chi = \sqrt{s}/(2 m_\chi)$ the rescaled energy 
of the two annihilating particles in their center-of-mass frame. The 
leading\footnote{We recall that we assume $m_f\ll T$, hence the 
neglected ${\cal O}(m_f^2 T^2/m_\chi^4)$ correction is sub-leading.} 
finite-temperature correction of order $T^4/m_\chi^4$ to the annihilation 
cross section is obtained by multiplying the scalar operator matrix elements 
with their Wilson coefficients (including the $\chi$-field part of the 
operator (\ref{eq:Oann})), and hence takes the form ($C_{f1} 
\equiv 0$)
\bea 
\label{eq:sigmavOPE2}
\sigma v_{\rm rel} &=&  \frac{2}{s}\sum_{n=1,2,3,4} 
\Big[\Im(C_{An}) M_{An} +  \Im(C_{fn}) M_{fn}\Big] \,.
\eea

The short-distance coefficients can be obtained by a standard 
zero-temperature calculation but depend on the specific model. To validate 
the OPE approach, we reconsider the model of \cite{Beneke:2014gla}, 
where the dark matter particle annihilates into a pair of fermions through the 
$t$-channel exchange of a scalar. In contrast to 
\cite{Beneke:2014gla}, we now assume that $\chi$ is a Dirac fermion, 
which avoids the P-wave suppression of the zero-temperature annihilation 
cross section to a pair of fermions. More precisely, 
we consider the process $\c\bar \c\to f^+ f^-$ in an
extension of the Standard Model by an $SU(2)\times U(1)$ 
singlet Dirac fermion $\c$ and a scalar doublet $\phi=(\phi^+,\phi^0)^T$. The 
relevant terms in the Lagrangian read
\begin{eqnarray}
\mathcal{L} &=& - \frac{1}{4} F^{\mu\nu} F_{\mu\nu} + \bar f 
\left(i\sl D-m_f \right) f 
+\bar\chi\left(i\sl \partial-m_\c \right)\chi\nonumber\\
&&  + \,
(D_\mu \phi)^\dagger (D^\mu \phi)-m_{\phi}^2 \phi^\dagger \phi +
\left(\lambda\bar\chi P_L f^- \phi^+ 
+\textrm{h.c.}\right),
\end{eqnarray}
where the SM fermions form a left-handed doublet $f=(f^0,f^-)^T$. 
The $t$-channel exchange of the heavy scalar $\phi$ occurs through 
a chiral Yukawa-type interaction.

We further assume that $m_\phi \gg m_\chi$ and integrate out the scalar 
mediator field in a first step. This leads to local 
annihilation operators $\mathcal{O}_{\rm ann} = J^\mu_t J_\mu$ of 
the general form of (\ref{eq:Oann}),
where we the ``current'' operators are given by
\bea
J_\mu & \equiv & \bar\chi P_L \gamma_\mu \chi \,,\\
J^\mu_t & \equiv &  \frac{\lambda^2}{2 m_\phi^2} \lc \lp 1 + 
\frac{m_\chi^2 + m_f^2}{m_\phi^2} \rp J_0^\mu + 
\frac{1}{m_\phi^2} J_1^\mu + \frac{1}{m_\phi^2} J_m^\mu \rc + 
\mathcal{O}(m_\phi^{-4})\,,
\label{eq:Jt}
\eea
with
\bea
J_0^\mu & \equiv & \bar f P_R \gamma^\mu f \,, \\[0.2cm]
J_1^\mu & \equiv & p_1^\alpha (iD_\alpha \bar f) P_R \gamma^\mu f + 
p_2^\alpha \bar f P_R \gamma^\mu (iD_\alpha f)\,, \\[0.2cm]
J_m^\mu & \equiv & -\frac{e}{4} F_{\alpha\beta} \lp  
\bar f \sigma^{\alpha\beta} P_R \gamma^\mu f +  
\bar f  P_R \gamma^\mu \sigma^{\alpha\beta} f   \rp
\eea
at tree level. Since the $\chi$ field is electrically neutral we have 
already simplified the expression by eliminating derivatives on $\chi$ 
($\bar \chi$) in favour of the momentum $p_1$ ($p_2$) of the 
particle, which literally arises only after taking the matrix 
element.

In the second step we construct the OPE (\ref{eq:OPE2}) of the above 
currents. We shall provide results explicitly only for the leading 
term in the expansion in $1/m_\phi^2$ in the first step, which allows 
us to simplify $J_t \to  \frac{\lambda^2}{2 m_\phi^2} J_0$. In this 
case, the calculation of the Wilson coefficients of the photon and 
fermion bilinear operators is very similar to the corresponding 
calculation in QCD, except that one cannot assume Lorentz invariance 
of the vacuum state, in which the operators are eventually evaluated.
The computation and the relevant diagrams are sketched in 
Appendix~\ref{sec:matching}. Defining $\widehat C_{i} = 2\,\Im (C_i)$
for the Wilson coefficients appearing in 
(\ref{eq:sigmavOPE2}), we find for the coefficients of the photon 
operators (using the notation of  \cite{Beneke:2014gla}, 
$\xi=m_\phi/m_\chi$, $\epsilon=m_f/2m_\chi$ and $\tau=T/m_\chi$):
\bea
m_\chi^{2} \,\widehat C_{A\, 1} & \propto & -6\epsilon^2 e_\chi^4 + 
42\epsilon^4 e_\chi^2 - 72\epsilon^6 , 
\label{eq:CA1} \\[0.2cm]
\widehat C_{A\, 2} = \widehat C_{A\, 4} & \propto & 
2 e_\chi^8 - 5\epsilon^2 e_\chi^4 \lp 4 e_\chi^2 - 1 \rp  + 
\epsilon^4 e_\chi^2 \lp 66 e_\chi^2 - 35 \rp - 12\epsilon^6 \lp 7 e_\chi^2 
- 5 \rp ,\quad  
\label{eq:CA2 } \\[0.2cm]
\widehat C_{A\, 3} & \propto & -10\epsilon^2 e_\chi^4 \lp 2 e_\chi^2 - 1 \rp  
+ 70\epsilon^4 e_\chi^2 \lp 2 e_\chi^2 - 1 \rp - 24\epsilon^6 \lp 11 e_\chi^2 
- 5 \rp , \quad  \label{eq:CA3}
\eea
where the proportionality factor reads
$\alpha\lambda ^4/(48 m_\chi^6 e_\chi^5 \xi^4 
\left(e_\chi^2-4 \epsilon ^2\right)^{5/2})$. 
We have kept here the full dependence on the fermion mass $m_f$, since 
for the contribution from the thermal photons our calculation is 
accurate also for $m_f ={\cal O}(T)$, and this provides a further 
check as discussed below. However, within the assumption 
$m_f\ll T\ll m_\chi$ adopted for the thermal fermion contribution, these 
terms can be dropped, in which case the coefficients simplify to 
\begin{equation}
\widehat C_{A\, 1} = \widehat C_{A\, 3} = 0\,,\qquad
\widehat C_{A\, 2} = \widehat C_{A\, 4} = 
\frac{\alpha\lambda ^4}{24m_\chi^6e_\chi^2\xi^4}\,.
\end{equation}
In case of the Wilson coefficients from the fermion operators, we set 
$m_f=0$ from the start (since we also neglected the condensates 
proportional to $m_f^2 T^2$ from the beginning), and obtain 
\be
\widehat C_{f2} = \widehat C_{f4} = 
\frac{\alpha \lambda^4}{12 m_\chi^6 e_\chi^2 \xi^4} \,,\qquad
\widehat C_{f3} =  0 \,. 
\label{eq:Cf24}
\ee

Assembling the coefficient functions and matrix elements according 
to (\ref{eq:sigmavOPE2}) we obtain the following three results:
\begin{itemize}
\item the tree-level, zero-temperature cross section, expanded in 
the heavy mediator mass,
\bea
\label{eq:sigmavtree_xiep_Dirac}
s\, \sigma_{\rm ann}^{\rm LO}\,v_{\rm rel}
&=&
\frac{\lambda^4 \sqrt{e_\chi^2-4\epsilon^2}}{24 \pi e_\chi\xi^4}\,
\Big( e_\chi^2 (4e_\chi^2 - 1) - 4\epsilon^2 (e_\chi^2 - 1) 
\Big) + \op(\xi^{-6})\,,
\eea
which follows from the coefficient function of the unit operator in 
the OPE (not given explicitly above). We also computed the next 
$\xi^{-6}$ term in the expansion by keeping the $1/m_\phi^2$ 
suppressed currents in the expression (\ref{eq:Jt}) for $J_t$.
\item the leading $\op(\tau^4)$ thermal correction to the annihilation 
cross section from the interactions with the photons in the 
plasma at order $\op(\xi^{-4})$:
\bea 
s\, \sigma_{\rm ann}v_{\rm rel}{}_{\,|\tau^4\, 
{\rm thermal\ photons}}&=&  
\frac{\pi^2\lambda^4 \alpha\tau^4}
{540\, e_\chi^3\, (e_\chi^2 - 4\epsilon^2)^{5/2} \, \xi^4}\, 
\Big( - e_\chi^6 (4 e_\chi^2 - 1)
\nonumber \\[0.2cm]
 && \hspace{-4cm}
 + \,2\epsilon^2 e_\chi^4 ( 22 e_\chi^2 - 7)
-\epsilon^4 e_\chi^2 (160 e_\chi^2 - 61) + 
 4\epsilon^6 (57 e_\chi^2 - 21) \Big) +\op(\xi^{-6})\,.\qquad
\label{eq:T4_phot_xi4_Dirac}
\eea
For vanishing final-state fermion mass, $m_f=0$, this expression 
simplifies to
\be
s\, \sigma_{\rm ann}v_{\rm rel}\,|^{\epsilon=0}_{\tau^4\, 
{\rm thermal\ photons}}
= -\frac{\pi^2 \lambda^4 \alpha \tau^4 \lp 4 e_\chi^2-1\rp  }
{540\, e_\chi^2 \,\xi ^4}\,.
\ee
\item the leading $\op(\tau^4)$ thermal correction to the annihilation 
cross section from the interactions with the fermions $f$ in the 
plasma at order $\op(\xi^{-4})$:
\be
\label{eq:T4_ferm_xi4_Dirac}
s\, \sigma_{\rm ann}v_{\rm rel}\,|^{\epsilon=0}_
{\tau^4,\, {\rm thermal\ fermions}} = 
\frac{7 \pi^2 \lambda^4 \alpha \tau^4 \lp 4 e_\chi^2-1\rp  }
{1080\, e_\chi^2 \,\xi ^4} \, ,
\ee
whose magnitude is 3.5 times larger than the thermal photon 
correction. 
\end{itemize}
In \cite{Beneke:2014gla} the corresponding and further results were 
obtained in the same model except for the Majorana nature of the 
dark matter particle, which involves further diagrams and leads 
to a suppressed zero-temperature cross section in the S-wave limit 
$e_\chi=1$. We have computed the Dirac case in the diagrammatic 
approach adopted in \cite{Beneke:2014gla} and confirmed the results 
given above.

As was the case for charged particle decay, the OPE computation
results in a significant conceptual and technical simplification. 
The infrared and collinear divergences present in individual 
diagrams in the diagrammatic approach are absent from the beginning. 
The OPE also provides direct insight 
into the temperature dependence of the thermal correction, and 
explains the absence of a ${\cal O}(T^2)$ correction for the 
neutral fermion dark matter annihilation observed in \cite{Beneke:2014gla} 
by the non-existence of a 
gauge-invariant operator of dimension~2. In general it provides 
a transparent interpretation of the thermal correction in terms 
of model-independent condensates in the thermal plasma and factorizes 
the model dependence in short-distance coefficients, which can be 
computed in conventional zero-temperature field theory.

%%%%%%%%%%%%%%%%%%%%%%%%%%%%%%%%%%%%%%%%%%%%%%%%%%%%%%%%%%%%%%%%%%%%%%%%%%%%

\section{Conclusions}
\label{sec:Conclusions}

In this paper we applied the operator product expansion technique 
to the decay and annihilation of heavy particles in a thermal medium 
with temperature below the heavy particle mass, $m_\chi$. This allows us to 
explain two interesting observations made before: a) that the leading 
thermal correction to the decay width of a charged particle is the 
same multiplicative factor of the zero-temperature width for 
a two-body decay and muon decay~\cite{Czarnecki:2011mr}, and b) that the 
leading thermal correction to fermionic dark matter annihilation arises 
only at order $T^4/m_\chi^4$~\cite{Beneke:2014gla}. The OPE further 
considerably simplifies the 
computation and factorizes it into model-independent matrix elements 
in the thermal background, and short-distance coefficients to be 
computed in zero-temperature field theory. 

While we considered here several specific examples, the method itself 
is general. An interesting extension, which combines the features of 
charged particle decay and neutral particle annihilation, concerns 
the annihilation of charged particles. In the context of dark matter 
relic density calculations this situation may arise if co-annihilations 
of dark matter particles with charged states are important. In this case 
it is expected from the above that the leading thermal correction 
appears already at  $\op(T^2/m_\chi^2)$ as in charged particle decay. 
However, since the correction is a combination of a typically 
weak interaction with the plasma and of $T/m_\chi$ suppression factors, 
it is generally still rather small.

\acknowledgments
This work was supported by the Gottfried Wilhelm Leibniz programme of the 
Deutsche Forschungsgemeinschaft (DFG) and the DFG cluster of excellence 
"Origin and Structure of the Universe." AH is supported by the University 
of Oslo through the Strategic Dark Matter Initiative (SDI).

%%%%%%%%%%%%%%%%%%%%%%%%%%%%%%%%%%%%%%%%%%%%%%%%%%%%%%%%%%%%%%%%%%%%%%%%%%%%
\appendix

\section{Calculation of the thermal vacuum elements}
\label{sec:AppA}

We begin with the matrix element of the second operator 
in~\eqref{eq:ops_with_indices},
\be
M_A^{\alpha\beta\gamma\delta} \equiv \bra{\Omega_T} \, 
F^{\alpha\beta}F^{\gamma\delta}\, \ket{\Omega_T}\, .
\ee
Since the only available four-vector is the plasma velocity $v$, 
antisymmetry of the field strength tensor and parity invariance of the 
electromagnetic interaction implies the general parametrization
\bea
\label{eq:M_Acoeffs}
M_A^{\alpha\beta\gamma\delta} =
c_1\lp g^{\alpha\gamma}g^{\beta\delta} - g^{\alpha\delta}g^{\beta\gamma} \rp
+ c_2 \lp v^{\alpha } 
\big(v^{\gamma } g^{\beta  \delta }-v^{\delta } g^{\beta  \gamma }\big) 
- v^{\beta } \big(v^{\gamma } g^{\alpha  \delta }
-v^{\delta } g^{\alpha  \gamma }\big) \rp. 
\quad
\eea
The coefficients $c_1$, $c_2$ are related to the contractions
\bea
\label{eq:TCeq1}
\bra{\Omega_T} \, F_{\mu\nu} F^{\mu\nu} \,\ket{\Omega_T} 
&=& 12 c_1 + 6 c_2 \, ,\\[0.1cm]
\label{eq:TCeq2}
\bra{\Omega_T} \, v_\mu v^\nu F_{\alpha\nu} F^{\alpha\mu} \,\ket{\Omega_T} 
&=& 3c_1 + 3 c_2 \,.  
\eea
Using the expression~(\ref{eq:Tpropgamma}) for the thermal photon 
propagator we find
\bea
\bra{\Omega_T}\, F^{\alpha\beta}F^{\gamma\delta}\,\ket{\Omega_T} 
&=& 
\int \frac{d^4 k}{(2\pi)^4}\,(-2\pi)\, \delta(k^2)\, f_B(\omega)
\nonumber \\[0.1cm]
&& \times \,\big( 
  k^\alpha k^\gamma g^{\beta\delta} 
- k^\alpha k^\delta g^{\beta\gamma} 
- k^\beta k^\gamma  g^{\alpha\delta} 
+ k^\beta k^\delta  g^{\alpha\gamma} \big)\, ,\qquad \quad
\eea
where $\omega = |v\cdot k|$ 
is the photon energy in the plasma frame. Performing the contractions 
and the integral results in 
\bea
\label{eq:TCsol1}
\bra{\Omega_T} \, F_{\mu\nu} F^{\mu\nu} \,\ket{\Omega_T} 
&=& 0 \,,
\\
\label{eq:TCsol2}
\bra{\Omega_T} \, v_\mu v^\nu F_{\alpha\nu} F^{\alpha\mu} \,\ket{\Omega_T} 
&=& -\frac{\pi^2}{15} T^4\,, 
\eea
in agreement with~\cite{Biondini:2013xua} (see also~\cite{CaronHuot:2009ns}),
and hence
\be
\label{eq:c123}
c_1 = -\frac{1}{2}\, c_2 = \frac{\pi^2}{45}T^4 \,. 
\ee

The matrix elements of the two fermion bilinear operators 
in~\eqref{eq:ops_with_indices} in the thermal vacuum  can be computed along 
the same lines. The thermal part of the fermion propagator is
\begin{equation}
\bra{\Omega_T}\, \timeord\!\left\{ f(x) \bar f(y) \right\} \ket{\Omega_T} 
= \int \frac{d^4k}{(2\pi)^4}\,e^{-ik\cdot(x-y)}
\, (-2\pi) \lp \slashed k + m_f \rp \delta(k^2-m_f^2)\, f_F(\omega)\,,
\end{equation}
with $f_F(\omega)$ the Fermi-Dirac distribution. Decomposing the general 
Dirac matrix $\Gamma$ into the basis $\left\{ \unitOP,\ \gamma_5,\ 
\gamma^\mu,\ \gamma^\mu\gamma_5,\ \sigma^{\mu\nu} \right\}\,$ 
we find that there are only two non-vanishing matrix elements of dimension 
four operators. There first involves $\Gamma=\unitOP$, in which case it 
follows immediately that $\bra{\Omega_T}\,m_f \bar f f \, \ket{\Omega_T} 
= {\cal O}(m_f^2 T^2)$ is suppressed for $m_f\ll T$. On the other hand, 
\begin{eqnarray}
M_f^{\mu\alpha} &\equiv& 
\bra{\Omega_T}\, \bar f\, \gamma^\mu iD^\alpha f\, \ket{\Omega_T} 
= 4 \int \frac{d^4 k}{(2\pi)^3} \, k^\mu k^\alpha \, 
\delta(k^2-m_f^2)\, f_F(\omega)
\nonumber \\
&=& d_1 \Big(\,g^{\mu\alpha} - 4 \,v^\mu v^\alpha\Big) + 
{\cal O}(m_f^2 T^2)\,,
\end{eqnarray}
where the relative coefficient of the two Lorentz tensors follows 
from the equation of motion $i\slashed D f=0$ in the $m_f\to 0$ limit.
The coefficient $d_1$ is found by multiplying with $v_\mu v_\alpha$, 
which yields 
\be
\label{eq:d123}
d_1 = - \frac{7\pi^2}{180}\,T^4 
\ee
up to $\op(m_f^2 T^2)$ corrections.

%%%%%%%%%%%%%%%%%%%%%%%%%%%%%%%%%%%%%%%%%%%%%%%%%%%%%%%%%%%%%%%%%%%%%%%%%%%%

\section{Calculation of the Wilson coefficients}
\label{sec:matching}

We briefly sketch the calculation of the coefficient functions of 
the operators (\ref{eq:O_Afinal}) in the operator production expansion 
(\ref{eq:OPE2}). As discussed in the main text, at ${\cal O}(\xi^{-4})$ 
we only need to consider the OPE of the product of two 
currents $J_0^\mu = \bar f P_R \gamma^\mu f$.

%%%%%%%%%%%%%%%%%%%%%%%%%%%%%%%%%%%%%%%%%%%%%%%%%%%%%%%%%%%%%%%%%%%%%%%%%%
\begin{figure}[t] 
 \centering
 \includegraphics[width=0.5\textwidth]{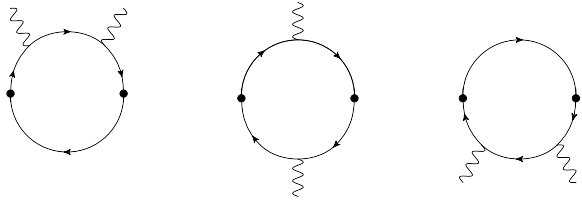}
\vskip0.0cm
 \caption{Diagrams contributing to the one-photon matrix element 
of the operator product $\op^{\mu\nu}(p)$. The dot denotes the 
insertion of the operator $J_0^\mu$, contracted with 
${\rm tr}\,\chi^{\mu\nu}$.}
 \label{fig:OPEdiag}
\end{figure}
%%%%%%%%%%%%%%%%%%%%%%%%%%%%%%%%%%%%%%%%%%%%%%%%%%%%%%%%%%%%%%%%%%%%%%%%%%

Referring to (\ref{eq:sigmav}) and (\ref{eq:sigmavOPE2}), the matching 
coefficients of the photon operators can be obtained by taking the 
one-photon matrix element. Denoting $\widehat C_{i} 
= 2\,\Im (C_i)$, the matching relation reads 
\be
\sum_{n = 1,\ldots,4} \widehat C_{A\,n}\,
\bra{\gamma}\, \op_{A\,n} \,\ket{\gamma} 
=  2\, {\rm Im}\,\big\{\bra{\gamma}\, \op^{\mu\nu}(p) \,\ket{\gamma}  \big\} 
\,{\rm tr}\,\chi_{\mu\nu}\,.
\label{eq:matchphoton}
\ee
Here 
\be
{\rm tr}\,\chi^{\mu\nu} = \frac{1}{4} \,
\sum_{\rm spin} \bra{\chi\bar\chi}\, 
(\bar\chi P_L \gamma^\mu \chi)( \bar \chi \gamma^{\nu} P_R \chi) \,
\ket{\chi\bar\chi} 
%\nonumber \\
= \frac{1}{4} \,\Tr \left\{ (\slashed p_2 - m_\chi)P_L 
\gamma^\mu (\slashed p_1 + m_\chi)\gamma^\nu P_R \right\} 
\quad
\ee
denotes the part that comes from contracting the fields in the 
dark matter current $J_\mu = \bar\chi P_L \gamma_\mu \chi$ with the 
external state, while $\op^{\mu\nu}(p)$ refers to the Fourier-transformed 
operator 
product on the left-hand side of (\ref{eq:OPE2}). The evaluation of 
the matrix element $\bra{\gamma}\, \op^{\mu\nu}(p) \,\ket{\gamma}$ 
involves the straightforward computation of the diagrams shown in 
Figure~\ref{fig:OPEdiag}, in an expansion in the small external 
photon momentum up to the second order, which yields 
(\ref{eq:CA1}) -- (\ref{eq:CA3}).

\begin{figure}[t] 
 \centering
 \includegraphics[width=0.75\textwidth]{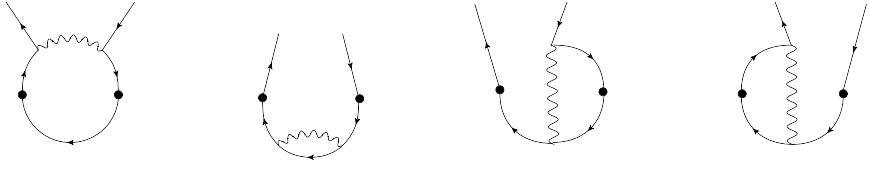}
\vskip0.0cm
 \caption{Diagrams contributing to the one-fermion matrix element 
of the operator product $\op^{\mu\nu}(p)$. The dot denotes the 
insertion of the operator $J_0^\mu$, contracted with 
${\rm tr}\,\chi^{\mu\nu}$.  The same diagrams with reversed charge flow 
are not shown explicitly.}
\label{fig:OPEdiags_ferm}
\end{figure}

Similarly, the matching equation for the fermion bilinear operators is 
\be
\sum_{n = 2,3,4} \widehat C_{fn} \,\bra{f}\, \op_{fn} \,\ket{f} 
=  2\, {\rm Im}\,\big\{ \bra{f}\, \op^{\mu\nu}(p) \,\ket{f}  \big\} 
\, {\rm tr}\,\chi_{\mu\nu}\,,
\ee
where the diagrams contributing to the one-fermion matrix element of 
the operator product are depicted in Figure~\ref{fig:OPEdiags_ferm}.
Note that diagrams without photon exchange do not contribute to 
the imaginary part at non-zero values of $s$ by momentum conservation. 
The calculation yields the result given in (\ref{eq:Cf24}).

%%%%%%%%%%%%%%%%%%%%%%%%%%%%%%%%%%%%%%%%%%%%%%%%%%%%%%%%%%%%%%%%%%%%%%%%%%
%%%%%%%%%%%%%%%%%%%%%%%%%%%%%%%%%%%%%%%%%%%%%%%%%%%%%%%%%%%%%%%%%%%%%%%%%%

%%%%%%%%%%%%%%%%%%%%%%%%%%%%%%%%%%%%%%%%%%%%%%%%%%%%%%%%%%%%%%%%%%%%%%%%%%
%\bibliography{bib_thermal}
%\bibliographystyle{JHEP-2}

\providecommand{\href}[2]{#2}\begingroup\raggedright\endgroup

\end{document}